# Effective Bandwidth Utilization in IEEE802.11 for VOIP

S.Vijay Bhanu
Research Scholar, Anna University, Coimbatore
Tamilnadu, India, Pincode-641013
E-Mail: vbhanu22@yahoo.in

Dr.RM.Chandrasekaran
Registrar, Anna University, Trichy
Tamilnadu, India, Pincode: 620024.
E-mail: aurmc@hotmail.com

Dr.V.Balakrishnan
Research Co-Supervisor
Anna University, Coimbatore
E-Mail :profdrvb@gmail.com

*Abstract -*Voice over Internet protocol (VoIP) is one of the most important applications for the IEEE 802.11 wireless local area networks (WLANs). For network planners who are deploying VoIP over WLANs, one of the important issues is the VoIP capacity. VoIP bandwidth consumption over a WAN is one of the most important factors to consider when building a VoIP infrastructure. Failure to account for VoIP bandwidth requirements will severely limit the reliability of a VoIP system and place a huge burden on the WAN infrastructure. Less bandwidth utilization is the key reasons for reduced number of channel accesses in VOIP. But in the QoS point of view the free bandwidth of atleast 1-5% will improve the voice quality. This proposal utilizes the maximum bandwidth by leaving 1-5% free bandwidth. A Bandwidth Data rate Moderation (BDM) algorithm has been proposed which correlates the data rate specified in IEEE802.11b with the free bandwidth. At each time BDM will calculate the bandwidth utilization before sending the packet to improve performance and voice quality of VoIP. The bandwidth calculation in BDM can be done by using Erlang and VOIP bandwidth calculator. Finally, ns2 experimental study shows the relationship between bandwidth utilization, free bandwidth and data rate. The paper concludes that marginal VoIP call rate has been increased by BDM algorithm.

## I. INTRODUCTION

VoIP services have been significantly gaining prominence over the last few years because of a number of impressive advantages over their traditional circuit-switched counterparts including but not limited to high bandwidth efficiency, low cost, and flexibility of using various compression strategies. In contrast to wired networks, the bandwidth of wireless network is limited. Furthermore, a wireless channel is error-prone and packets can be discarded in transmission due to wireless errors such as signal fading or interference. Thus, the efficiency of a wireless channel access becomes a critical issue.

Currently, the most popular WLAN standard is the IEEE 802.11b, which can theoretically support data rates up to 11 Mb/s, however, this data rate is for optimal conditions [1]. On the other hand, 802.11a and 802.11g networks have data rates up to 54 Mb/s and they are not designed to support voice transmission (because of the APs are not distributed in the most optimum way, communication can be established properly) [2]; they are used for data transmission, and a network only designed for data transmission is not ideal for voice transmission. Compare to data packet, voice packets are small in size. Due to the large overhead involved in transmitting small packets, the bandwidth available for VoIP traffic is far less than the bandwidth available for data traffic. This overhead comprises transmitting the extra bytes from various networking layers (packet headers) and the extra time (backoff and deferral time) imposed by the Distributed Coordination Function (DCF) of 802.11b.

This paper experimentally study the relationship between bandwidth utilization in the wireless LAN and the quality of VoIP calls transmitted over the wireless medium. On an 802.11 b WLAN, frames are transmitted at up to 11 Mbps. There is a lot of overhead before and after the actual transmission of frame data, however, and the real maximum end-to-end throughput is more on the order of 5 Mbps. So, in theory, 802.11b should be able to support 50-plus simultaneous phone calls[1]. But practically it support only 5 calls. This proposal improves bandwidth utilization in order to achieve maximum channel access and improved QoS by using BDM algorithm. The number of channel access can be improved by changing the data rate frequently.

This paper is structured as follows: Section IA describes about basic history of 802.11 MAC and previous related work. Section III introduces a method for predicting VoIP bandwidth utilization. Section IV shows the BDM algorithm and its functionalities and Section V&VI discuss about the simulation topology, parameters and results. Final part contains conclusion and future enhancement.

### A. Basic Theory of 802.11 MAC

The basic 802.11 MAC protocol is the Distributed Coordination Function (DCF), which is based on the Carrier Sense Multiple Access/Collision Avoidance (CSMA/CA) mechanism [3] [4]. A mobile station (STA) is allowed to send packets after the medium is sensed idle for the duration greater than a Distributed Inter-Frame Space (DIFS). If during anytime in between the medium is sensed busy, a back-off procedure should be invoked. Specifically, a random variable uniformly distributed between zero and a Contention Window (CW) value should be chosen to set a Back-off Timer. This Back-off Timer will start to decrement in units of slot time, provided that no medium activity is indicated during that particular slot-time. The back-off procedure shall be suspended anytime the medium is determined to be busy and will be resumed after the medium is determined to be idle for another DIFS period. The STA is allowed to start transmission as soon as the Back-off Timer reaches zero. A mobile station (STA) shall wait for an ACK when a frame is sent out. If the ACK is not successfully received within a specific ACK timeout period, the STA shall invoke back-off and retransmission procedure. The CW value shall be increased exponentially from a CWmin value until up to a CWmax value during each retransmission.

An additional Request to Send/ Clear To Send (RTS/CTS) mechanism is defined to solve a hidden terminal problem inherent in Wireless LAN. The successful exchange of RTS/CTS ensures that channel has been reserved for the transmission from the particular sender to the particular receiver. The use of RTS/CTS is more helpful when the actual data size is larger compared with the size of RTS/CTS. When the data size is comparable with the size of RTS/CTS, the overhead caused by the RTS/CTS would compromise the overall performance.

## II. PREVIOUS WORKS

This section, reviews the existing literature related to enhancing voip call capacity. In reference [5] Aggregation with fragment Retransmission (AFR) scheme, multiple packets are aggregated into and transmitted in a single large frame. If errors happen during the transmission, only the corrupted fragments of the large frame are retransmitted. Clearly, new data and ACK frame formats are a primary concern in developing a practical AFR scheme. Optimal frame and fragment sizes are calculated using this model, and an algorithm for dividing packets into near-optimal fragments is designed. Difficulties for new formats include 1) respecting the constraints on overhead noted previously and 2) ensuring that, in an erroneous transmission, the receiver is able to retrieve the correctly transmitted fragments—this is not straightforward because the sizes of the corrupted fragments may be unknown to the receiver.

Extended dual queue scheme (EDQ) provides a QoS for the VoIP service enhancement over 802.11 WLAN. It proposes a simple software upgrade based solution, called an Extended Dual queue Scheme (EDQ), to provide QoS to real-time services such as VoIP [6]. The extended dual queue scheme operates on top of the legacy MAC. The dual queue approach is to implement two queues, called VoIP queue and data queue. Especially, these queues are implemented above the 802.11 MAC controllers, i.e., in the device driver of the 802.11 network interface card (NIC), such that a packet scheduling can be performed in the driver level. Packets from the higher layer or from the wire line port (in case of the AP) are classified to transmit into VoIP or data types. Packets in the queues are served by a simple strict priority queuing so that the data queue is never served as long as the VoIP queue is not empty. But the hardware upgrade is undesirable.

The cross-layer scheme of [7] [8] is named as Vertical Aggregation (VA) since it works along the same flow. The main advantage is that it enhances voice capacity using a plain IEEE802.11 MAC protocol, and adopting an additional application aware module, logically placed above the MAC layer. In reference [9] proposes two feedback-based bandwidth allocation algorithms exploiting HCCA to provide service with guaranteed bounded delays: (1) the Feedback Based Dynamic Scheduler (FBDS) and (2) the Proportional Integral (PI)-FBDS. They have been designed using classic discrete-time feedback control theory. We will assume that both algorithms, running at the HC, allocate the WLAN channel bandwidth to wireless stations hosting real-time applications, using HCCA functionalities. This allows the HC to assign TXOPs (transmission opportunity) to ACs by taking into account their specific time constraints and transmission queue levels. We will refer to a WLAN system made of an Access Point and a set of quality of service enabled mobile stations (QSTAs). Each QSTA has up to 4 queues, one for each AC in the 802.11e proposal. FBDS require a high computational overhead at the beginning of each service period, due to the queue length estimation.

By [8] Wireless Timed Token Protocol (WTTP) provides traffic streams with a minimum reserved rate, as required by the standard, and it accounts for two types of traffic streams simultaneously, depending on the corresponding application: constant bit rate, which are served according to their rate, and variable bit rate traffic streams. Additionally, WTTP shares the capacity which is not reserved for QoS traffic streams transmissions among traffic flows with no specific QoS requirements. This VAD [10] algorithm is capable of removing white noise as well as frequency selective nose and maintaining a good quality of speech.

## III. CALCULATING BANDWIDTH CONSUMPTION FOR VOIP

Bandwidth is defined as the ability to transfer data (such as a VoIP telephone call) from one point to another in a fixed amount of time. The bandwidth needed for VoIP transmission will depends on a few factors: the compression technology, packet overhead, network protocol used and whether silence suppression is used. Voice streams are first encapsulated into RTP packets, and they are carried by UDP/IP protocol stack [3]. A single voice call consists of two opposite RTP/UDP flows. One is originated from the AP to a wireless station, and the other oppositely flows. There are two primary strategies for improving IP network performance for voice: several techniques were proposed for QoS provisioning in wireless networks[11] [12]. Allocate more VoIP bandwidth and implement QoS.

How much bandwidth to allocate depends on:

- Packet size for voice (10 to 320 bytes of digital voice)

- CODEC and compression technique (G.711, G.729, G.723.1, G.722, proprietary)

- Header compression (RTP + UDP + IP), which is optional

- Layer 2 protocols, such as point-to-point protocol (PPP), Frame Relay and Ethernet

- Silence suppression / voice activity detection

Calculating the bandwidth for a VoIP call is not difficult once you know the method and the factors to include. The chart below, "Calculating one-way voice bandwidth," demonstrates the overhead calculation for 20 and 40 byte compressed voice (G.729) being transmitted over a Frame Relay WAN connection [13]. Twenty bytes of G.729 compressed voice is equal to 20 ms of a word.

Voice digitization and compression:

G .711: 64,000 bps or 8000 bytes per second

G.729: 8000 bps or 1000 bytes per second

Protocol packet overhead:

IP = 20 bytes, UDP = 8 bytes, RTP =12 bytes

Total:40 bytes

If one packet carries the voice samples representing 20 milliseconds, the 50 such samples are required to be transmitted in every second. Each sample carries an IP/UDP/RTP header overhead of 320 bits [14]. Therefore, in each second, 16,000 header bits are sent. As a general rule of 'thumb', it can be assumed that header information will add 16kbps to the bandwidth requirement for voice over IP. For example, if an 8kbps algorithm such as G.729 is used, the total bandwidth required to transmit each voice channel would be 24kbps.

The voice transmission requirements are,

- Bandwidth requirements reduced with compression, G.711, G.729 etc.

- Bandwidth requirements reduced when longer packets are used, thereby reducing overhead.

- Even though the voice compression is an 8 to 1 ratio, the bandwidth reduction is about 3 or 4 to 1. The overhead negates some of the voice compression bandwidth savings.

- Compressing the RTP, UDP and IP headers is most valuable when the packet also carries compressed voice.

**A**. Packet Overhead

To support voice over WLANs, it is important to reduce the overhead and improve the transmission efficiency over the radio link. Recently, various header compression techniques for VoIP have been proposed [14]. The RTP/UDP/IP headers can be compressed to as small as 2 bytes.

## IV. PROPOSED ALGORITHM

The main reason for the extra bandwidth usage is IP and UDP headers. VoIP sends small packets and so many times, the headers are actually much larger than the data part of the packet. The proposed algorithm based on the following two factors. 1) A small frame is in error then there is a high probability of error for a large frame as well. Similarly when a large frame is successful, there is a very high probability of success for small frames as well. 2) The amount of free bandwidth decreases as the number of VoIP calls increases. As well as the call quality decreases as the number of VoIP calls increases. Free Bandwidth (BWfree) that corresponds to the remaining unused idle time that can be viewed as spare or available capacity. In BDM algorithm, at each frame transmission will calculates the free bandwidth availability.

Variables

BWfree: Unused idle bandwidth viewed as spare or available capacity
BWload: Specifies the bandwidth used for transmission of the data frames
Drate: It specifies the data rate
Incr: Increment operation
Decr: Decrement operation

Functions:
UpperLevel (): upper level according to table 1, 2
LowerLevel (): lower level according to the table 1, 2

A. BDM ALGORITHM:
Initial level:
Drate: LowerLevel ()
BWfree: UpperLevel ()
S: Previous transmission
If (S = Success)
{
Incr Drate to next UpperLevel ()
Decr BWfree to next LowerLevel ()
}
Else
{
Decr Drate to next LowerLevel ()
Incr BWfree to next UpperLevel ()
}

According to IEEE 802.11b only four types of data rates are available, which are 1, 2, 5.5, 11mbps. When the data rate is high then the throughput increases at the same time the chance for occurring error also increases [1] [15]. To avoid this situation BDM allocates some free bandwidth to improve the QoS. This free bandwidth allocation should be at the minimum level otherwise again quality degradation occurs.

TABLE1: LEVELS OF DATA RATE

| Levels | Data Rate |
|--------|-----------|
| Level 0 | 1 mbps |
| Level 1 | 2 mbps |
| Level 2 | 5.5 mbps |
| Level 3 | 11 mbps |

TABLE2: LEVELS OF BANDWIDTH FREE

| Levels | % of Free Bandwidth |
|--------|---------------------|
| Level 0 | 1 |
| Level 1 | 2 |
| Level 2 | 3 |
| Level 3 | 4 |
| Level 4 | 5 |

Number of calls = Correc_Fac ( RB-RBT ) / Codec
Where,
Correc_Fac: Correction factor of real network performance
RB: Real bandwidth usage
RBT: Real bandwidth used for data transmission
Codec: Bandwidth used by the codec to establish a call

End-to-end (phone-to-phone) delay needs to be limited. The shorter the packet creation delay, the more network delay the VoIP call can tolerate. Shorter packets cause less of a problem if the packet is lost. Short packets require more bandwidth, however, because of increased packet overhead (this is discussed below). Longer packets that contain more speech bytes reduce the bandwidth requirements but produce a longer construction delay and are harder to fix if lost. By BDM the data rate and free bandwidth will improve the number of VOIP calls as well as performance.

## V. SIMULATION TOPOLOGY

The simulation study is conducted using the ns-2 simulator. The simulation result will be compared with IEEE 802.11 specifications. Any node can communicate with any other node through base station. The number of stations can be varied from 5 to 50. Wireless LAN networks are set up to provide wireless connectivity within a finite coverage area of 20 to 30m. The network simulator will be used to form an appropriate network topology under the Media Access Control (MAC) layer of the IEEE 802.11b. According to the IEEE 802.11b protocol specifications [16], the parameters for the WLAN are shown in Table 3. When calculating bandwidth, one can't assume that every channel is used all the time. Normal conversation includes a lot of silence, which often means no packets are sent at all. So even if one voice call sets up two 64 Kbit RTP streams over UDP over IP over Ethernet (which adds overhead), the full bandwidth is not used at all times.
Based on [2] the data rate and coverage area will be changed. 802.11b standard can cover up to 82 meters of distance, considering that only the first 48 meters are usable for voice, the other 34 meters are not usable, therefore cellular area must be fit only to the 48 meters from the AP in order to avoid interferences, which is depicted in table 4.

TABLE 3: PARAMETERS USED FOR SIMULATION

| Parameter | Value |
|-----------|-------|
| DIFS | 50 $\mu$sec |
| SIFS | 10 $\mu$sec |
| Slot time | 20 $\mu$sec |
| CWmin | 32 |
| CWmax | 1023 |
| Data Rate | 1,2,5.5,11 Mbps |
| Basic rate | 1 Mbps |
| PHY header | 192 $\mu$sec |
| MAC header | 34 bytes |
| ACK | 248 $\mu$sec |

TABLE4: DATA RATES AND DISTANCE FOR VOIP

| Data rate in Mbps | Distance in meters |
|-------------------|--------------------|
| 54 | 0-27 |
| 48 | 27-29 |
| 36 | 29-30 |
| 24 | 30-42 |
| 18 | 42-54 |
| 11 | 0 - 48 |

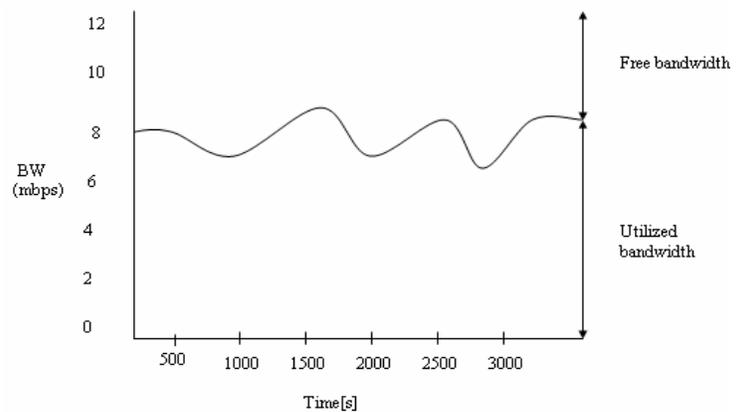

Fig 1: Free bandwidth analysis

Free bandwidth= Total bandwidth-bandwidth utilized
          =100-87.5
          =13.5

In this sample calculation the free bandwidth is 13.5%. From this 8.5% of bandwidth can be utilized for frame transmission to achieve maximum throughput and leave 5% to obtain Qos. Fig 1 shows the difference between bandwidth utilization and free bandwidth. When the amount of free bandwidth dropped below 1% call quality became unacceptable for all ongoing calls. The amount of free bandwidth is a good indicator for predicting VoIP call quality, but in the throughput point of view it should be reduced. This contradiction can be solved by using BDM algorithm.

## VI. SIMULATION RESULTS

### A. Throughput

The throughput (measured in bps) corresponds to the amount of data in bits that is transmitted over the channel per unit time. In the following Fig 2 X-axis specifies timeslot and Y-axis specifies the throughput. Consider for each time slot the channel receives 10 frames. When the time slot is 4ms, the throughput is 4200kbps, when it increases to 8ms it is 7000kbps. The graph shows the gradual improvement and the overall throughput is increased upto 87.5%.

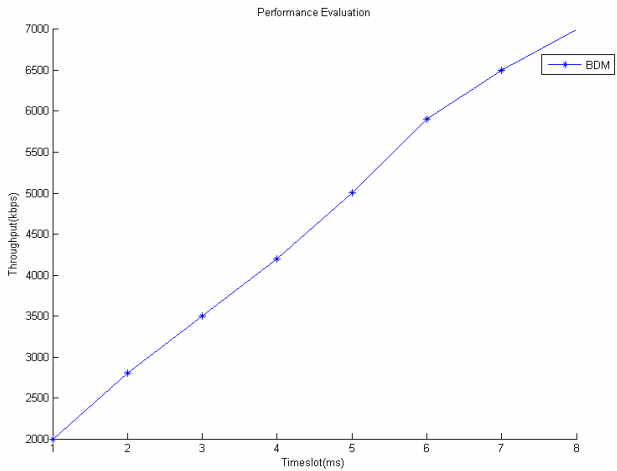

Fig 2: Variations in throughput with respect to timeslot

### B. Frame Loss

Frame loss is expressed as a ratio of the number of frames lost to the total number of frames transmitted. Frame loss results when frames send are not received at the final destination. Frames that are totally lost or dropped are rare in WLANs. In the fig-3 X-axis shows the number of frames and Y-axis shows the frame loss percentage. The frame loss value increased upto 0.5 when it reaches a threshold value it slowly decreasing . These frame loss degradation will improve our VOIP performance in a great manner.

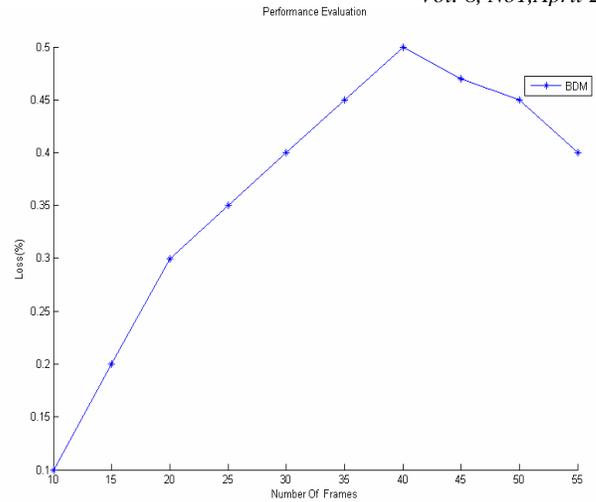

Fig 3: Variations in packet loss when number of frames increases

### C. Delay

Investigating our third metric, average access delay for high priority traffic, Fig 4 shows that has very low delays in most cases, even though the delays increases when the load gets very high. However, all the schemes have acceptable delays [6], even though EDCA in most cases incur a longer delay than the other schemes. Even if a scheme can give low average access delay to high priority traffic, there might still be many packets that get rather high delays. With the number of data stations increasing, the delay performance of the voice stations degrades. This tells that the VoIP performance is sensitive to the data traffic.

In fig-4 graph X-axis specifies the number of frames and Y-axis specifies the delay in ms. When the number of frames is in between 5-10 the delay is gradually increased after that there is no change in the delay. Nearly 30% of the delay is reduced by BDM algorithm.

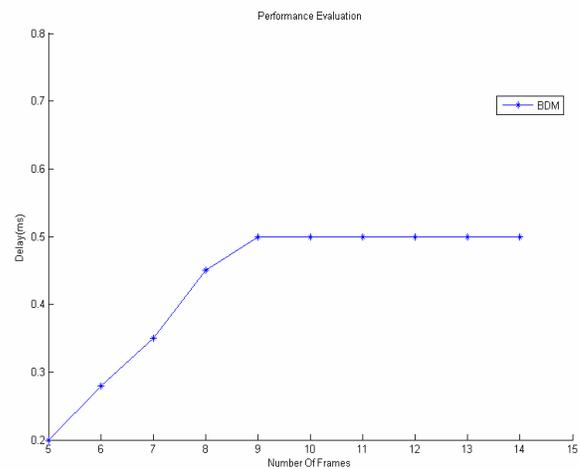

Fig 4: Variations in delay when number of frames increases

## D. Bandwidth Utilization

In most cases, the normal VoIP telephone call will use up 90Kbps. When calculating bandwidth, one can't assume that every channel is used all the time. Normal conversation includes a lot of silence, which often means no packets are sent at all. So even if one voice call sets up two 64 Kbit RTP streams over UDP over IP over Ethernet (which adds overhead), the full bandwidth is not used at all times. Note that for the graphs where the priority load is low, the utilization first increases linearly.

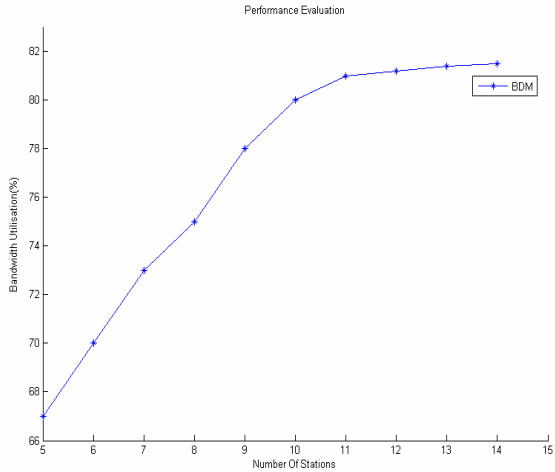

Fig5: Variations in bandwidth utilization

In fig-5 X-axis specifies the stations and Y-axis specifies the bandwidth utilization percentage. From the number of stations 5 the curve starts incrementing, when the stations become 15 the bandwidth utilization percentage is beyond 80%. In the above specified graph 30% of overall bandwidth utilization is increased.

## E. Throughput Comparison

Here the throughput performance of the EDCA algorithm and our proposed BDM algorithms are compared [17]. By using values of maximum achievable throughput from simulation, VoIP capacity in WLAN can also be evaluated. The following formula is used for getting the average packets sent from AP and all VoIP nodes in one second.

Capacity = Maximum Throughput / Data Rate

In BDM algorithm the data rate is changed frequently in an effective manner, So that the overall capacity will be improved. When data rate increases, automatically the throughput will increase. Due to the low level of data transfer rate the throughput seldom reach 600 kbps in EDCA. Due to moderate data rate the maximum throughput in BDM is 1000 kbps, it shows that 15% improvement in overall throughput. From the graph, the EDCA algorithm can support only 7.5 calls but BDM algorithm can support 9.1 voip calls (only for

10 stations). This calculation shows that 16% of overall voip call rate is increased by BDM algorithm.

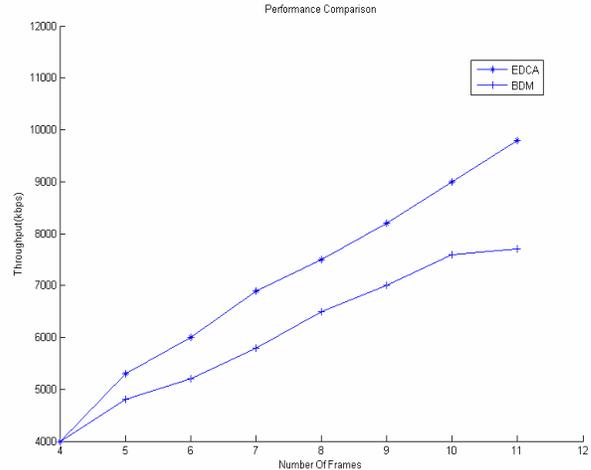

Fig 6: Variations in throughput when number of frames increases

## F. Comparison of Bandwidth Utilization

VoIP bandwidth consumption over a WAN (wide area network) is one of the most important factors to consider When building a VoIP infrastructure [9]. Failure to account for VoIP bandwidth requirements will severely limit the reliability of a VoIP system and place a huge burden on the WAN infrastructure. Short packets require more bandwidth, however, because of increased packet overhead (this is discussed below). Longer packets that contain more speech bytes reduce the bandwidth requirements but produce a longer construction delay and are harder to fix if lost.

In Fig.7 X-axis shows the number of stations and Y-axis shows the bandwidth utilization percentage. When the number of frame is 25 the EDCA algorithm gives only 65% of bandwidth utilization and it start to decrease if the number of stations exceeds 40. But BDM algorithm gives 85% of bandwidth utilization also the curve is gradually increases when the number of stations increased. By BDM algorithm 20% of bandwidth utilization is increased.

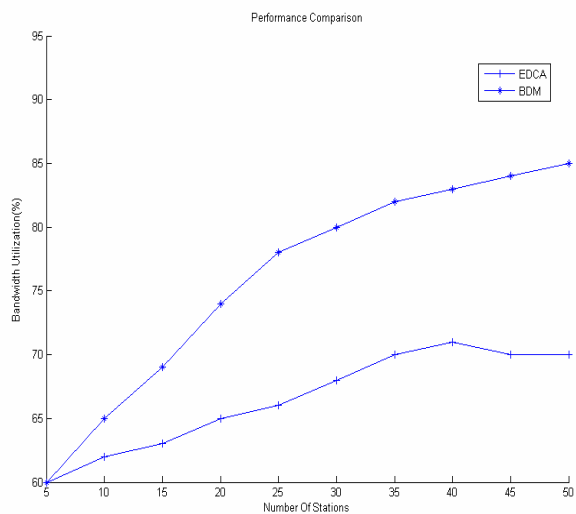

Fig 7: Variations in bandwidth utilization

## VII. CONCLUSION

This proposal discusses a new association algorithm called bandwidth and data rate moderation algorithm for IEEE 802.11 stations. BDM is designed to enhance the performance of an individual station by calculating the free bandwidth availability. When several users are working simultaneously, the real bandwidth is divided among the whole users. Through experimentation with a number of VoIP calls and various data rates in an 802.11b WLAN shoes a close relationship between wireless bandwidth utilization and call quality. When the amount of free bandwidth dropped below 1% call quality become unacceptable for all ongoing calls. Simulated result shows that marginal improvement in VOIP call rate is realized by BDM algorithm. The future work will consider the different type of codec techniques and coverage area to increase the bandwidth utilization.

AUTHORS PROFILE

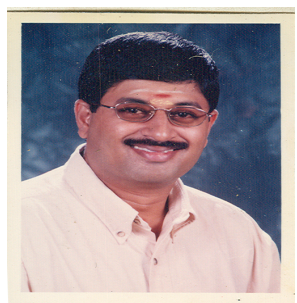

Mr. **S. Vijay Bhanu** is currently working as Lecturer (senior Scale) in the Computer Science & Engineering Wing, Directorate of Distance Education, Annamalai University. He is an co-author for a monograph on Multimedia. He served as wing Head, DDE, Annamalai University, Chidambaram for nearly five years. He served as Additional Controller of Examination at Bharathiar University, Coimbatore for two years. He conducted a workshop on Business intelligence in the year 2004. He is a life Member in Indian Society for Technical Education. Email: vbhanu22@yahoo.in

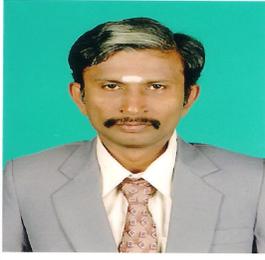

**Dr.RM.Chandrasekaran** is currently working as a Registrar of Anna University, Tiruchirappalli and Professor at the Department of Computer Science and Engineering, Annamalai University, Annamalai Nagar, Tamilnadu, India. From 1999 to 2001 he worked as a software consultant in Etiam, Inc, California, USA. He received his Ph.D degree in 2006 from Annamalai University, Chidambaram. He has conducted workshops and conferences in the area of Multimedia, Business Intelligence, Analysis of Algorithms and Data Mining. Ha has presented and published more than 32 papers in conferences and journals and is the co-author of the book Numerical Methods with C++ Program( PHI,2005). His research interests include Data Mining, Algorithms and Mobile Computing. He is life member of the Computer Society of India, Indian Society for Technical Education, Institute of Engineers and Indian Science Congress Assciation. Email: aurmc@hotmail.com

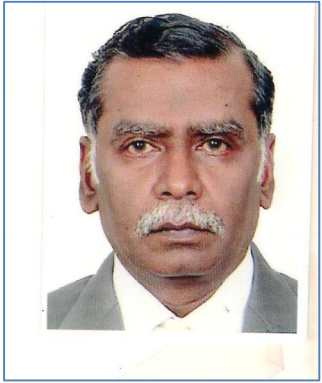

Dr.V.Balakrishnan, formerly Director, Anna University, Coimbatore has got 35 years of service to his credit in teaching, research, training, extension, consultancy and administration. He has guided 30 M.Phil Scholars, guiding 15 Ph.D Scholars. He guided around 1000 MBA projects. He got 13 international and national awards including, 'Management Dhronocharaya' and 'Manithaneya Mamani'. At Anna University he introduced 26 branches in MBA. He published around 82 articles in international and national Journals. He partook in about 100 international and national seminars. He served in various academic bodies like Academic Council, Faculty of Arts, Board of Selection, and Board of Examiners in most of the Universities in South India.